# Design recommendations based on speech analysis for disability-friendly interfaces for the control of a home automation environment

Nadine Vigouroux[1], Frédéric Vella[1], Gaëlle Lepage[2,3], Éric Campo[4]

[1]IRIT, UNRS 5505, Université Paul Sabatier, 118 Route de Narbonne, 31062 Toulouse, France
[2]GIHP 10 Rue Jean Gilles, 31100 Toulouse, France
[3]UT2J, Campus Mirail, 5 allée Antonio Machado 31058 Toulouse, France
[4]LAAS, CNRS, UT2J, 7 Avenue du Colonel Roche, 31400 Toulouse, France

mailto:nadine.vigouroux@irit.fr,frederic.vella@irit.fr,
gaelle1704@gmail.com,eric.campo@laas.fr

**Abstract.** The objective of this paper is to describe the study on speech interaction mode for home automation control of equipment by impaired people for an inclusive housing. The study is related to the HIP HOPE project concerning a building of 19 inclusive housing units. 7 participants with different types of disabilities were invited to carry out use cases using voice and touch control. Only the results obtained on the voice interaction mode through the Amazon voice assistant are reported here. The results show, according to the type of handicap, the success rates in the speech recognition of the command emitted on the equipment and highlight the errors related to the formulation, the noisy environment, the intelligible speech, the speech segmentation and the bad synchronization of the audio channel opening.

**Keywords:** spoken interaction, speech disorder, motor impairment, visually impaired and hearing impairment, smart home.

## 1    Introduction

In France, the ELAN law of November 23, 2018 introduces the concept of inclusive housing and defines it as a mode of housing (excerpt from Article L.281-1 of the CASF): "*intended for people with disabilities and the elderly who have the choice, as their main residence, of a grouped mode of living, among themselves or with other people (…) and accompanied by a social and shared life project*". Due to the changes and obstacles of this new type of housing, it seems interesting to study the technological needs of people with disabilities, taking into account their physical and material environment, allowing them to live in an inclusive habitat with the greatest possible autonomy.

Varriale et al. [1] have identified the role and function of home automation, for people with disability through a deep review, specifically they aim to outline if and how the home automation solutions can support people with disability improving their social inclusion. Vacher et al. [2] described an audio-based interaction technology that allows the user to have full control over their home environment and detect distress situations for the elderly and visually impaired people. These two research works show the need to study home automation technologies for the autonomy of people with disabilities.

The study presented in this paper is related to the HIP HOPE project concerning a building on the Montaudran site in Toulouse (France) in which 19 inclusive housing units will be built. This inclusive housing project is a participatory housing project characterized by economic, generational, social and cultural diversity. It will welcome students, as well as and disabled people, families with young children and seniors. This project aims at social inclusion through culture, popular education and solidarity. It concerns people with all types of disabilities, including a majority of residents with motor or neurological disabilities, or with neurodegenerative diseases (multiple sclerosis, Charcot's disease, Huntington's disease, Parkinson's disease), ...

These homes are part of an architectural and environmental project based on the principles of sustainable development and energy savings, in a warm, friendly, open, bright and secure housing approach. The project will include up to twenty apartments within a larger residence (100 to 120 units) in order to allow for a mix of publics and social inclusion. The cultural component will promote this social inclusion. Each apartment of the project will be designed to be accessible and will be equipped with home automation with simple and adapted control functionalities such as voice and touch interaction, and not only by push button type. Vigouroux et al. [3] reported the experiment design carried out in a living Lab to identify the most accessible interaction modes for home automation controls.

In this context, the objective of this article is to describe an experiment conducted in this living lab to focus on the voice interaction mode for home automation control of equipment by impaired people in order to provide recommendations for its implementation in HIP HOPE housing.

Section 2 addresses the various researches in the field of speech recognition for the elderly and people with motor and speech impairments. Section 3 describes the experimental framework with emphasis on the infrastructure of the Smart Home of Blagnac (MIB)[1], the study population and the scenarios. Section 4 analyses the interaction logs and proposes a typology of speech recognition errors. The last section discusses the results and proposes research perspectives for a better usability and acceptability of speech recognition and voice assistants in the context of smart homes.

## 2    Related works

Advances in voice recognition and natural language processing have reached a level of maturity and performance that their deployment in smart houses, on smartphones etc. is becoming widespread. Today, people are using this technology to perform a lot of everyday tasks in their homes like home automation (turn off/on lights, turn off/on TV, etc.).

---

[1] http://mi.iut-blagnac.fr

The literature review shows that recent studies explore the usability and the usefulness of automatic speech recognition and voice assistants for elderly people [4] and for people with motor and speech impairment [5]. Recently Brewer et al. [6] reported a work focused on perceptions of voice assistant use by elderly people. The authors showed how conversational and human-like expectations with voice assistants lead to information breakdowns between the older adult and voice assistant. They also discussed how voice interfaces could better support older adults' health information seeking behaviours and expectations.

Kowalski et al. [7] explored the possible benefits and barriers to the use of Google Home voice assistants in the context of smart home technology by older adults. They found that older adults could naturally identify various already available applications of voice assistants such information hand-free, translator or memory aid, etc. However, the group of older adults reported that with little training and encouragement older participants could start using voice assistants to their satisfaction and empowerment.

We can observe many designs of voice applications for the elderly and people with motor disabilities integrating voice assistants [8], [9], [10] and [11]. Some applications use voice components such as Google Home [7], Google Assistant [12] and Amazon's Alexa [13].

Alexakis et al. [8] described an IoT Agent, a Web application for monitoring and controlling a smart home remotely. This IoT Agent integrates a chat bot that can understand text or voice commands using natural language processing. This current implementation is totally based on third party APIs and open source technologies. Their implementation exploits the typical IoT infrastructures by providing an enhanced user experience and usability of the underlying IoT system with the integration of natural language processing and voice recognition, MQTT and many other technologies. Although the article describes the implementation in depth, the authors make no mention of performance measurements on the language processing tools.

Ismail et al. [9] proposed low cost framework on smartphone and Raspberry Pi boards which uses a hybrid Support Vector Machine (SVM) with a Dynamic Time Warping (DTW) algorithm to enhance the speech recognition process. The results show 97% accuracy and that patients and elderly people are accessing and controlling IoT devices in smart homes and hospitals. The system's limitation is the difficulty of speech recognition if the user's voice is affected by illness or is not clear enough to be detected.

Lokitha et al. [10] proposed a comparative study of two algorithms: the first is the Support Vector Machine (SVM) and the second is Convolutional Neural network (CNN) for command identification of Speech disabled and Paralyzed People. The CNN model yields an accuracy of 90.62%, whereas the SVM algorithm gives a very low accuracy (58.42%). The authors suggest increasing the learning database with real speech signals collected from patients to improve the performance of the system future and of the model for continuous speech.

Netinant et al. [11] have designed a speech recognition system on raspberry Pi using the Google Speech Recognition API with a special emphasis on disabled and elderly individuals. However, the authors reported that their system's accuracy in recognizing speech for four command words in Thai and English language for disabled and elderly users should be weighed against the country's national language. While this study is interesting, it is limited in the number of speakers and orders.

Isyanto et al. [12] implemented IoT-based smart home voice commands for disabled people using Google Assistant to control TV, lights, etc. However, the authors reported that, the Google Assistant application accepts voice commands when the pronunciation is correct.

Mtshali and Khubisa [14] presented a system smart objects and a voice assistant such as Amazon Alexa, Google Home, Google Assistant, Apple Siri, etc. to recognize voice commands, from a person with physical disabilities through three use cases. However, the authors do not mention any evaluation of the accuracy obtained.

Malavasi et al. [15] have developed low-cost voice systems for environmental control through simplified and accessible user interfaces by adapting to dysarthric speech, or to more general language disorders through adaptation processes.

This related background shows that many research and innovation approaches are underway in terms of recognition methods and tools, in the use of OpenHAB interoperability tools [15], MQTT[2] (Message Queuing Telemetry Transport) communication protocols [8], https (HyperText Transfer Protocol), and some studies on the acceptability of voice assistants, mainly by the elderly.

We also note that these systems are becoming widespread on smartphones and embedded boards such as Raspberry Pi. Although all the studies agree on the benefits of this mode of interaction for autonomy of elderly and disabled people, several authors highlight the need to acquire speech disorder databases in order to build acoustic models adapted to these disorders and the need to conduct experiments on larger populations with speech disorders.

## 3 Experiment

### 3.1 Material

**Living Lab description**

The experimentation was carried out in MIB which served as a Living Lab. Its objective is to provide researchers with a study and research platform to observe people in situation using the different prototypes. It has an apartment (see **Fig. 1**) of 70m² allowing making experiments with impaired people. It is composed of different rooms: living room, kitchen, corridor, bedroom, bathroom and toilet. It is equipped with various connected objects and motorized furniture that can be controlled by speech interaction, touch on tablets/phones or by wall switches such as a removable sink and washbasin in the kitchen and toilet, television, a medical bed, shutters, lighting and automatic alert devices. It also has an infrastructure for experiments with microphones, cameras and motion sensors.

---

[2] https://mqtt.org/

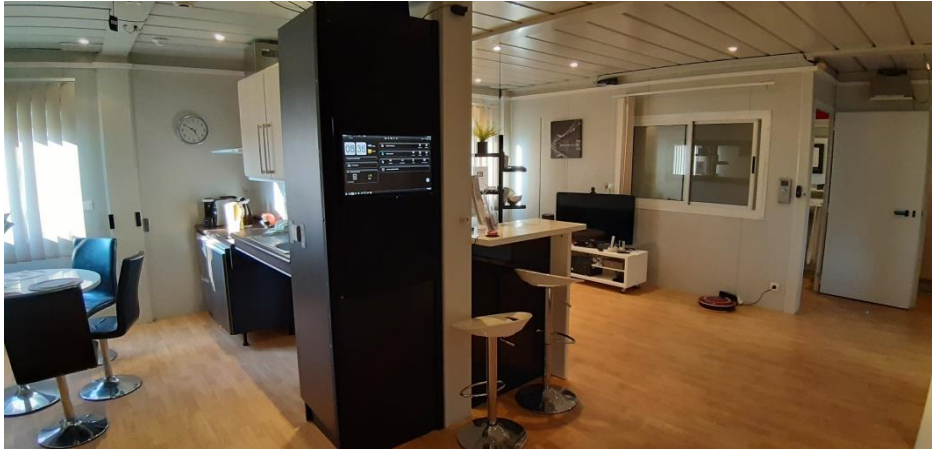

**Fig. 1.** Smart Home at the University Institute of Technology in Blagnac.

**Infrastructure description**

The MIB infrastructure is composed of an MQTT Broker server, an OpenHAB interoperability server, a touch device, a Fire TV Cube voice assistant from Amazon and several connected objects that are connected to a Wifi box (see **Fig. 2**) which allows to assign an IP address to each connected object.

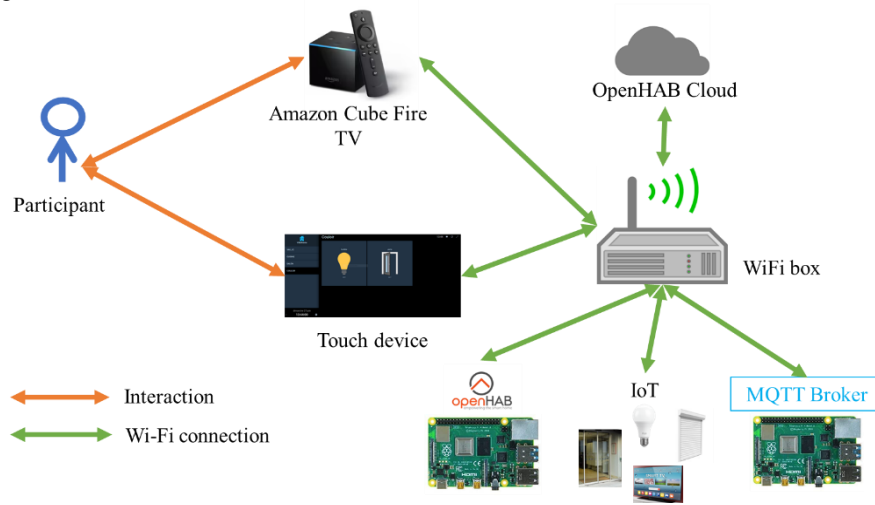

**Fig. 2.** Overall system architecture.

We selected the interoperability platform because it allows the interconnection of connected objects using communication protocols (MQTT, https, etc.). It also enables the design of interfaces by means of tools that it supports (for example, the HABPanel application for the design of graphic interfaces). It includes some skills to connect the voice assistants such as Amazon Echo, Google Home, etc.

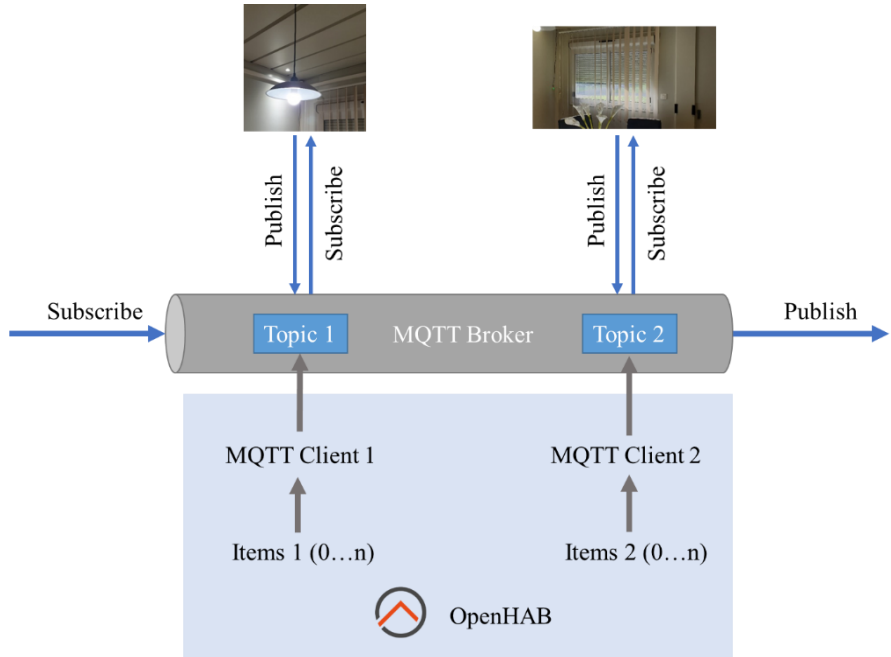

**Fig. 3.** Connection between OpenHAB and MQTT.

We then chose MQTT as the communication protocol (see **Fig. 3**) to connect Internet Of Things to interaction buttons (for instance, to turn on/off the light, to open/close the shutters, etc.). To include the connected objects on OpenHAB, we first defined the MQTT topics of the different objects in the house in the platform. We then defined the OpenHAB items of these objects by giving them item labels and interaction modes (two-state switches –on versus off– three-state switches –up, stop, down –, etc.). In addition to these items, we can associate the "Amazon" metadata that will allow Amazon to access these items for the recognition process.

For example, for the bedroom light we first have the MQTT ((Message Queuing Telemetry Transport) topics described in a hierarchical way [16] to:

- request the light status: api/1/room/bedroom/lamp/ceiling/id/1/indication (*publish*)
- turn the light on and off: api/1/room/bedroom/lamp/ceiling/id/1/request followed by on or off (*subscribe*).

The OpenHAB platform also allows two items to be distinguished for the same MQTT topics. This functionality allows the design of several interaction modes, for example:

- one for the touch tablet called "touch room light"
- and another for the voice assistant called "voice room light" by adding the "Amazon" metadata.

This metadata allows the Amazon Cube Fire TV device to access item labels stored on the OpenHAB Cloud to connect Internet Of Things with the voice recognition (see **Fig. 2**). Indeed, the OpenHAB platform secures the access of items through the Cloud by means of credentials. To configure Amazon's Fire TV Cube, it is necessary to download the Alexa application on a smartphone and then integrate the OpenHAB skill from the Alexa application.

We can then associate the action "open the shutters" in Fire TV Cube with several language utterances such as "*open the shutters of the living room*", "*open living room shutters*", "*open the roller shutters in the living room*", etc. All these utterances should be associated with the "living room shutter" item in OpenHAB.

OpenHAB also allows the design of more advanced use cases consisting of elementary actions as described above. We use deterministic rules with respect to the item states and the item order to design use cases. Based on this principle, we have developed the call "for help" which consists of the following elementary actions: I open the front door; I turn on the hall light; I turn on the bedroom light and a spoken message "help is on the way" is broadcast throughout the house.

**Voice assistant**

For the speech interface, we chose the Fire TV Cube from Amazon because it allowed controlling by voice the television with a Molotov account and the home automation.

The list of items defined in OpenHab illustrates how the light in the bedroom is switched on.

```
- Item 'GF_Bedroom_Light' received command ON
- Item 'GF_Bedroom_Light' predicted to become ON
- Item 'GF_Bedroom_Light' changed from OFF to ON
```

### 3.2 Population

The participants were recruited through an association for the integration of physically disabled people (GIHP). This set of participants of all ages and with different impairments represents the population that will live in the HIP HOPE home automation flats. They are representative of the residents' profile who will occupy these inclusive habitats: 7 persons with impairments (1 intellectual impairment, 4 motor disabled people including 2 with speech disorders, 1 visual and 1 hearing impairments) participated in this study. Table 1 also lists the home automation and assistive technologies desired by these individuals, collected from interviews.

Table 1. Table of participants extracted from [3].

| Participant | Age/Gender | Impairment | Activities | Technology needs for smart home |
|---|---|---|---|---|
| 101 | 63/M | Hearing impairment | Pharmacist, now retired | Adapted intercom with high quality visuals to see the person and read their lips; connected objects with visual feedback; flashing lights; app on phone to detect someone's presence or an abnormal noise. |
| 102 | 72/M | Visual impairment | Computer science now retired | Easy to implement; efficient and responsive technology, limit the number of steps, preference for voice control with voice feedback on actions performed; home automation control (shutters, light, alarm) but with reliability and ease of use. |
| 104 | 39/F | Cerebral Palsy | Employee in an association and volunteer | Interfaces for home automation control (shutters, front door); voice control difficult in case of fatigue, so have the touch mode; connected intercom without the need to pick up the phone. |
| 202 | 18/M | Trisomy syndrome | Student | Smartphone application to help organise activities, to encourage initiatives (coaching application). |
| 204 | 19/M | Cerebral palsy | Student | Smartphone control system for gates, garages and front doors to be autonomous; smartphone remote control for TV, robotic arm. |
| 300 | 38/F | Myopathy | Volunteer | Home automation to control the environment (with voice command); robotic arm (help for cutting, grabbing objects, grooming), adapted intercom (easy to open and to communicate). |
| 302 | 70/F | Polio | Secretary, retired and volunteer | Opening of the gate from your home; automated bay window; automation control of equipment for individual and mobile homes; fall detector or easy emergency call. |

### 3.3 Scenarios

We defined the scenarios (see **Table 2** and **Table 3**) based on the environmental control needs expressed during the interviews. We invited the participants to use the speech interaction modality described (Section 3.1) to control the equipment of the Smart Home in two use cases (a controlled and a free scenario).

Table 2. Controlled scenario: "Controlling your environment".

| | |
|---|---|
| Instruction: | Use the touch tablet and/or the voice command to control the house. |
| Initial conditions of the MIB: | Shutters closed, all lights on, motorized furniture is in down position. The person can be with their caregiver in the house but they must not interact with the voice and touch controls. |
| Example of a control: You go into the kitchen | Open the shutter, Turn off the light, Raise the kitchen furniture, etc. |

Table 3. Free scenario: "Activities of daily living when getting out of bed"

| | |
|---|---|
| Instruction: | Use the touch tablet and/or the voice command to control the house. |
| Initial conditions of the MIB | The participant is sitting on the bed or in the wheelchair with the lights off. When the scenario is started, the participant is asked to perform a number of tasks in the order he/she wishes. |
| Example of tasks | You go and have breakfast; You turn on the TV in the living room to watch a news channel, sitting on the sofa or in your armchair; You feel unwell and call for help (panic button, medallion or voice command) "*Alexa, help*". |

We have invited the participants to perform both the free and the controlled scenario. However, they were free to choose the order of using the tactile modality or the spoken modality. Then, we gave the scenario either in paper or in digital form for ease of understanding, depending on the participant's impairment. The visual impaired participant had the scripts printed in Braille.

**Fig. 4** shows on the bottom left screen the visualization of the participant's speech signal, on the top right screen the view of the four rooms of the house and the status of the connected objects in the house. These screens enabled the experimenter to observe the participant's behavior.

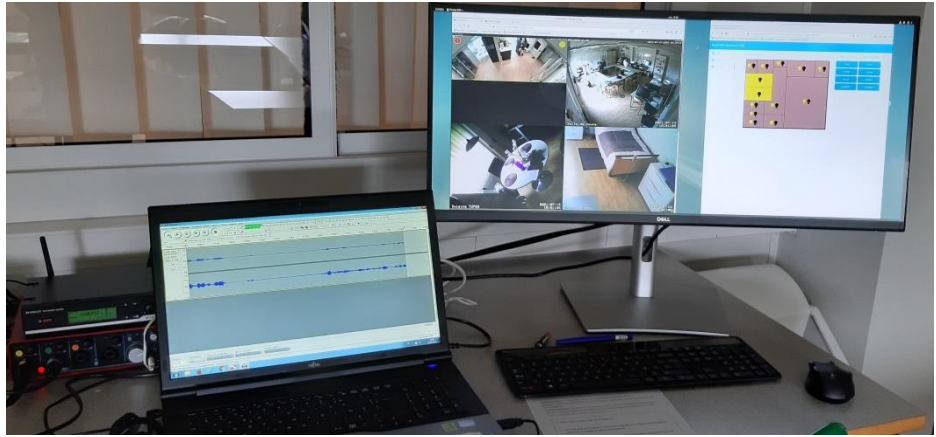

**Fig. 4.** Experimenter's observation workstation.

### 3.4 Appropriation phase

The experimenter proceeded to demonstrate the two interaction modes for the same command. He then explained the need to open the recognition audio channel of the Amazon Fire TV system. The experimenter drew the participant's attention to the need for feedback from the Amazon Fire TV system (voice feedback "okay") and visually (blue light display). Given the language difficulties of some participants, we selected the channel opening word that was easiest for the participant to pronounce, from these three words (Amazon, Alexa, Echo). The experimenter then invited the participant to carry out three types of command in each mode to ensure that the instructions were correctly understood. The duration of this task varied greatly depending on the participants, mainly for participants with speech disorders.

### 3.5 Speech Interaction Data

All dialogues between the participant and the Amazon Fire TV were transcribed in order to match the channel corresponding to the utterance produced by the participant with the one recognized by the recognition system. We also transcribed the sentence corresponding to the channel openings (See Table 4).
The aim of this transcription is to measure the recognition performance but also the difficulties of opening the communication channel of this type of voice assistant for our study population. We have added a comment field in which the transcriber can notify, for example, the bad synchronisation of the production of the utterance with the opening of the channel, syllable or word accentuations according to the participants, pauses between the entities of a syntactic structure, etc.

**Table 4.** Transcription format.

| Type of speech act | Spoken utterance | Recognized utterance | Comments |
|---|---|---|---|

## 4 Results

The results are deduced from the analysis of the transcribed data in the format described in Table 4.

### 4.1 Spoken interaction rate

Participants produced 146 utterances with a very variable number of utterances depending on the participant (4 utterances for participant 204 including 3 reformulations and 47 utterances for participant 104). The overall correct recognition rate was 76.03% and the recognition error rate was 23.97%. The **Fig. 5** shows that participant 104 had a recognition error rate of 53.2% and a reformulation rate of 10.6%. The experimenter always opened the communication channel for this participant. Participant 204 had a recognition error rate of 100%. This participant did not wish to continue with this mode of interaction. For participant 300, the recognition rate is 100%. Participants 101, 202, 302 and 102 had an average recognition error of 12.13% and a standard deviation of ±2.96. With the exception of participants 104 and 204, no other participant had a problem opening the communication channel.

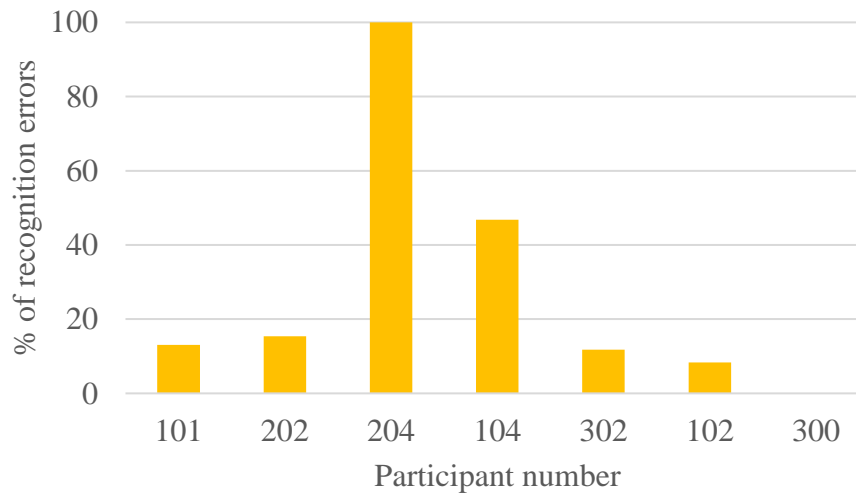

**Fig. 5.** Pourcentage of speech recognition per participant

### 4.2 Typology of recognition errors

We have carried out a typology of recognition errors in order to propose recommendations for the design of voice assistants and voice interaction systems for people impairments. The **Fig. 6** illustrates the percentage according to the causes of errors.

- Communication channel opening problem

We identified two reasons: 1) The speaker did not speak the trigger keyword (Alexa, Amazon, Echo) or did not speak loud enough so the communication channel was not opened but the speaker continued to speak. In this case, the oral utterance was not subject to the recognition process; 2) A synchronization problem between the recognition process and the speaker. We have identified two situations: either the person started speaking immediately before the feedback from the Fire Cube TV was obtained, or he/she spoke too late after the feedback. In both cases, the utterance was not subjected to the recognition process.

- Noisy environment

The environment was noisy (loud TV) and the system did not recognize the utterance.

- Utterance segmentation

These errors corresponded to pauses made between words in an utterance (participant 202 and 204) and or the accentuation of syllables at the end of a word. This mode of speech corresponded to speech in isolated words.

- Unintelligible speech

This speech was also unintelligible to a human being. The speech reading and the interaction context (location of the person in the room) did not enable the experimenter to understand the communication intention of the participant. Participant 204 wished to abandon voice interaction despite efforts to rephrase in favor of tactile interaction.

- Recognition errors

Errors that could not be classified in the previous categories are considered as recognition errors without classification. They generally corresponded to phonetic confusions (mispronunciation of the word "living room" in the statement "open the shutter of the living room") by the recognition system due to the participant's speech difficulties.

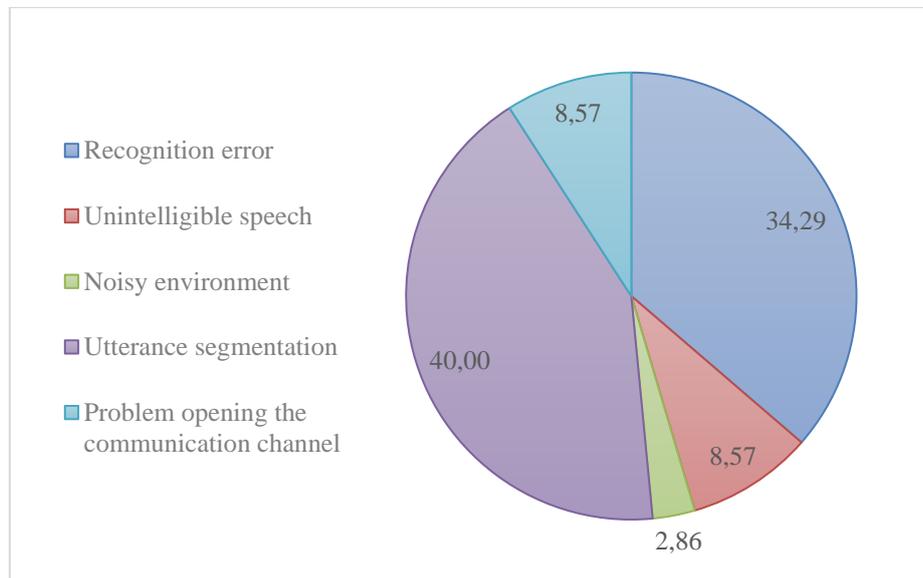

**Fig. 6.** Typologies of speech recognition errors in %.

## 5 Discussion

This pre-study shows that recognition accuracy is variable depending on the speakers' communication abilities. We have identified two difficulties that speech technologies will have to take into account, namely:

- Articulatory difficulties in the production of oral utterances due to speech disorders. Analyzes of the participants' oral utterances (204 and 104) identified the following difficulties: the production of words in syllables, the accentuation of some phonetic units, pauses between words, repetitions, and a slower speech rate. Thus, the acoustic representation of the speech signal of speakers with speech disorders is different from those without speech disorders. We propose some recommendations to adapt recognition systems to speech disorders:
    - Record databases with various speech disorders to adapt the acoustic models of recognition units like Lokitha et al. [10];
    - Adapt the language models to take into account syntactic structures (pause, repetitions, syllabification of words, etc.);
    - Link simplified utterances to the same command to avoid production errors as much as possible (e.g., instead of the utterance "open the living room roller shutters", accept also the utterance "open living room shutter"); This simplification of grammar was proposed by Malavasi et al. [15];
    - Use other multiple information such as lip movement to increase the likelihood of the recognition system as done by the hearing impaired.

- Their ability to pronounce in the right tempo (the opening word of the communication channel and the oral statement corresponding to the command). The bad management of the communication channel (not waiting for the feedback of Amazon, forgetting to open the channel) is a blocking point. The results of this pre-study suggest that the appropriation phases of the vocal interaction are more important, especially for subjects with attentional difficulties (participant 202), to adjust the intervals of maintaining the opening channel according to the participants. For participant 104, the experimenter opened the communication channel. To remedy this situation, we propose a push button or a button on an interface (tablet, smartphone) where the validation of a click would emit a pre-recorded vocal message that would allow the opening of the communication channel.

5 of the 7 participants preferred speech interaction to tactile interaction. Participant 204 would prefer to use voice if the speech recognition worked for him "because with fatigue it would be more efficient". People with speech disorders do not want to be excluded from voice interaction.

## 6    Conclusion

The development of the smart home facilitates human activities. However, the problem is the limited activities for disabled people. This paper describes an experiment with seven different disabled people interacting with smart home equipment with the Amazon Alexa assistant.

The related work shows that many technological innovations on speech recognition for smart homes are being carried out. This maturity of speech recognition technology has enabled usability and acceptability studies for elderly and disabled people. Secondly, the infrastructure based on OpenHAB and MQTT is described. Then, we report the speech accuracy for each participant with an average accuracy of 76.03 %. We note very poor performance (100 % and 46.8 % of false recognition rate) for the two speech impaired participants. This suggests two research axis to be pursued: the development of acoustic and language models for people with speech impairments, the development of more intelligent voice assistants integrating more developed natural language and dialogue components, the proposal of alternative modalities such as interactive interaction or multimodal combination (voice, tactile, lip movement, eye movement, gesture). In terms of perspective, we propose to extend our study population to non-technophile people and to analyze tactile interactions.

## Acknowledgment

The study is partially funded by the Occitanie Region (France). The authors thank the participants and the GIHP association.

# References

1. Varriale, L., Briganti, P., & Mele, S. (2020). Disability and Home Automation: Insights and Challenges within Organizational Settings. In Exploring Digital Ecosystems, Springer, 47–66. DOI: 10.1007/978-3-030-23665-6_5
2. Vacher, M., Lecouteux, B., Istrate, D., Joubert, T., Portet, F., Sehili, M., & Chahuara, P. (2013). Experimental evaluation of speech recognition technologies for voice-based home automation control in a smart home. In 4th workshop on Speech and Language Processing for Assistive Technologies, 99-105. https://hal.archives-ouvertes.fr/hal-00953244
3. Vigouroux, N., Vella, F., Lepage, G., Campo, E. (2022). Usability Study of Tactile and Voice Interaction Modes by People with Disabilities for Home Automation Controls. In: Miesenberger, K., Kouroupetroglou, G., Mavrou, K., Manduchi, R., Covarrubias Rodriguez, M., Penáz, P. (eds) Computers Helping People with Special Needs. ICCHP-AAATE 2022. Lecture Notes in Computer Science, vol 13342. Springer, Cham. https://doi.org/10.1007/978-3-031-08645-8_17
4. Vacher, M., Caffiau, S., Portet, F., Meillon, B., Roux, C., Elias, E., Lecouteux, B., Chahuara, P. (2015). Evaluation of a Context-Aware Voice Interface for Ambient Assisted Living: Qualitative User Study vs. Quantitative System Evaluation. ACM Trans. Access. Comput. 7, 2, Article 5, 36 pages. https://doi.org/10.1145/2738047
5. Jefferson, M. (2019). Usability of automatic speech recognition systems for individuals with speech disorders: Past, present, future, and a proposed model
6. Brewer, R., Pierce, C., Upadhyay, P., & Park, L. (2022). An empirical study of older adult's voice assistant use for health information seeking. ACM Transactions on Interactive Intelligent Systems (TiiS), 12(2), 1-32.
7. Kowalski, J., Jaskulska, A., Skorupska, K., Abramczuk, K., Biele, C., Kopeć, W., & Marasek, K. (2019). Older adults and voice interaction: A pilot study with google home. In Extended Abstracts of the 2019 CHI Conference on human factors in computing systems,1-6
8. Alexakis, G., Panagiotakis, S., Fragkakis, A., Markakis, E., & Vassilakis, K. (2019). Control of smart home operations using natural language processing, voice recognition and IoT technologies in a multitier architecture. Designs, 3(3), 32.
9. Ismail, A., Abdlerazek, S., & El-Henawy, I. M. (2020). Development of smart healthcare system based on speech recognition using support vector machine and dynamic time warping. Sustainability, 12(6), 2403.
10. Lokitha, T., Iswarya, R., Archana, A., Kumar, A., Sasikala, S. (2022). Smart voice assistance for speech disabled and paralyzed people. In 2022 International Conference on Computer Communication and Informatics (ICCCI), IEEE, 1-5
11. Netinant, P., Arpabusayapan, K., & Rukhiran, M. (2022). Speech Recognition for Light Control on Raspberry Pi Using Python Programming. In 2022 The 5th International Conference on Software Engineering and Information Management (ICSIM), IEEE, 33-37
12. Isyanto, H., Arifin, A. S., & Suryanegara, M. (2020). Design and implementation of IoT-based smart home voice commands for disabled people using Google Assistant. In 2020 International Conference on Smart Technology and Applications (ICoSTA), IEEE, 1-6
13. Yue, C. Z., & Ping, S. (2017). Voice activated smart home design and implementation. In 2017 2nd International Conference on Frontiers of Sensors Technologies (ICFST) 489-492
14. Mtshali, P., & Khubisa, F. (2019). A smart home appliance control system for physically disabled people. In 2019 Conference on Information Communications Technology and Society (ICTAS), IEEE, 1-5
15. Malavasi, M., Turri, E., Motolese, M. R., Marxer, R., Farwer, J., Christensen, H., ... & Green, P. (2017). An innovative speech-based interface to control AAL and IoT solutions to help people with speech and motor disability. In Ambient Assisted Living: Italian Forum 2016 7 (pp. 269-278). Springer International Publishing.
16. Van den Bossche, A., Gonzalez, N., Val, T., Brulin, D., Vella, F., Vigouroux, N., & Campo, E. (2018). Specifying an MQTT Tree for a Connected Smart Home. In: International Conference On Smart homes and health Telematics (ICOST 2018), Singapore. DOI: 10.1007/978-3-319-94523-1\_21.